\documentclass[aps,pre,twocolumn,showpacs,superscriptaddress,groupedaddress]{revtex4-2}

\usepackage{amsmath}
\usepackage{amsfonts}
\usepackage{amssymb}
\usepackage{graphicx}
\usepackage{graphics}
\usepackage{dcolumn}
\usepackage{sidecap}
\usepackage{parskip}
\usepackage{xcolor}
\usepackage{hyperref}
\usepackage{hhline}
\usepackage{mathtools}
\usepackage{multirow}
\usepackage{verbatim}
\usepackage{rotating}
\usepackage{setspace}
\usepackage{epsfig}
\usepackage{epstopdf}
\usepackage{natbib}
\usepackage{subfigure}
\usepackage{booktabs}
\usepackage[normalem]{ulem}
\usepackage{bm}

\usepackage{verbatim}
\usepackage{comment}
\usepackage{cleveref}
\usepackage[normalem]{ulem}

\makeatletter
\def\@eqnnum{{\normalsize \normalcolor (\theequation)}}
 \makeatother

\graphicspath{{./}{ER/main/}}
\usepackage{tikz} 

\begin{document}

\title{Stochastic Kuramoto oscillators with inertia and higher-order interactions}
\author{Priyanka Rajwani}
\email{phd2201151017@iiti.ac.in}
\author{Sarika Jalan}
\email{sarika@iiti.ac.in: Corresponding Author} \affiliation{Complex Systems Lab, Department of Physics, Indian Institute of
Technology Indore, Khandwa Road, Simrol, Indore-453552, India}

\begin{abstract}
Impact of noise in coupled oscillators with pairwise interactions has been extensively explored. 
Here, we study stochastic second-order coupled Kuramoto oscillators with higher-order interactions, and show that as noise strength increases the  critical points associated with synchronization transitions shift toward higher coupling values. 
By employing the perturbation analysis, we obtain an expression for the forward critical point as a function of inertia and noise strength. Further, for overdamped systems we show that as noise strength increases, the first-order transition switches to second-order even for higher-order couplings. We include a discussion on nature of critical points obtained through Ott-Antonsen ansatz.
\end{abstract}
    
\maketitle
\paragraph{\bf{Introduction:}} Synchronization is a fundamental phenomenon observed across physics, biology, chemistry, and engineering; from the rhythmic flashing of fireflies, the coordinated firing of neurons in the brain, and the stability of power grids \cite{osipov2007}. The Kuramoto model provides a pivotal analytical tool for studying synchronization \cite{kuramoto1975, strogatz2000}, and demonstrating how a system of interacting oscillators with diverse natural frequencies begins moving in unison as the interaction strength varies. This model is particularly valued for its simplicity and analytical tractability, which not only make theoretical analyses feasible but also enhance its utility in practical applications such as power system \cite{guo2021} and biological systems \cite{bick2020}. Additionally, research has explored how the adaptation function and phase lag parameter affect the synchronized state \cite{khanra2020, omelchenko2012, manoranjani2023, biswas2024}. 
Introducing noise to the Kuramoto model incorporates stochasticity, which reflects the intrinsic fluctuations found in real-world systems. Thus, this approach enables more accurate and realistic simulations, effectively mimicking real-world environments \cite{hindes2023}. 
Campa and Gupta considered the case of heterogeneous noise and analyzed the impact of noise strength on dynamical evolution of  coupled Kuramoto oscillators \cite{campa2023}. There exist few other studies investigating the impact of noise strength on synchronization profile in networks of phase oscillators \cite{esfahani2012, hindes2023}, and on the model used to study stability of large human connectome graph \cite{odor2021}. 

Furthermore, incorporation of inertia in the coupled Kuramoto system provides an application for modeling power grids \cite{filatrella2008, tanaka1997, rohden2012}. An inclusion of inertia term in the Kuramoto model has been shown to lead first-order phase transitions characterized by abrupt changes in system dynamics in response to a small change in coupling strength \cite{gao2018}.  Managing external perturbations or noise is crucial to prevent systemic failures in many complex systems \cite{olmi2014}. Acebr\'{o}n {\it{et. al.}} studied the influence of noise on critical transition points at which systems dynamics exhibit significant changes \cite{acebron1998, acebron2000}. Gupta {\it{et. al.} }outlined a detailed phase space diagram for coupled Kuramoto oscillators with pairwise interactions \cite{gupta2014, komarov2014}. Additionally, Cao {\it{et. al.}} described the effects of noise on cluster explosive synchronization in second-order Kuramoto oscillators on networks \cite{cao2018}.

All these results were limited to Kuramoto model (with or without inertia) with pairwise interactions. Recent studies have emphasized importance of higher-order interactions in modeling real-world complex systems
 \cite{battiston2020networks, boccaletti2023, gao2023dynamics}. Incorporating higher-order interactions into the Kuramoto model has been shown to lead abrupt (de)synchronization transition \cite{skardal2020, jalan2022} and tiered synchronization \cite{skardal2022, rajwani2023} in contrast to second-order transitions observed in pairwise interactions. In absence of noise, the second-order Kuramoto model with higher-order interactions revealed the presence of prolonged hysteresis  \cite{sabhahit2024}. Furthermore, studies on the second-order Kuramoto model with higher-order interactions, incorporating phase lag and coupling strengths ranging from negative to positive, have demonstrated the emergence of synchronization and frequency chimera states \cite{jaros2023}.
 
 This Letter considers the second-order Kuramoto model with higher-order interactions and Gaussian white noise. The study investigates noise-induced transitions in the model. Using Fokker-Plank equation, we first derive distribution function for frequency and phases for identical oscillators, and observe that the frequency order parameter is modulated by inertia and remains unaffected by coupling strength of both pairwise and higher-order interactions. Second, we show that an increase in noise strength shifts the forward (backward) critical point associated with an abrupt jump from an incoherent to coherent state (vice versa) toward higher coupling values. By employing a perturbation analysis, we obtain the expression for critical coupling strength in the forward direction which shows a dependence on inertia and noise strength. 
 Additionally, in an overdamped system, we analytically predict all (un)stable states using the Ott-Antonsen approach. Also, in presence of higher-order interactions, note a shift from first-order to second-order phase transitions as noise strength increases. 
 
\paragraph{\bf{Model:}}
We consider a stochastic Kuramoto model with $2-$simplex interactions and inertia. The equation of motion of $N$ globally coupled oscillators is given as,
\begin{equation}\label{Model_Eq}
\begin{split}
    m\ddot{\theta_i}=&-\dot{\theta_i}+{\Omega_i}+\frac{K_1}{N}\sum_{j=1}^{N}\sin(\theta_j-\theta_i)\\ &+\frac{K_2}{N^2}\sum_{j=1}^{N}\sum_{k=1}^{N}\sin(2\theta_j-\theta_k-\theta_i)+\xi_i(t),
\end{split}
\end{equation}
where $\theta_i$ and $\Omega_i$ indicate the phase and intrinsic frequency, respectively, of $i^{th}$ Kuramoto oscillator. The parameters $K_1$ and $K_2$ denote the coupling strength for pairwise and $2-$simplex interactions, respectively, and $m$ is the inertia term. The noise term $\xi_i$ is defined as white Gaussian noise, with $\langle\xi_i(t)\rangle=0$ and $\langle\xi_i(t)\xi_j(s)\rangle=2D\delta_{ij}\delta(t-s)$, where $D$ represents the noise strength. The order parameter is defined as:
\begin{equation} \label{ord_par}
    {r_pe^{\iota\psi_p}=\frac{1}{N}\sum_{j=1}^{N}e^{\iota p \theta_j}},
\end{equation}
\DeclareRobustCommand{\inlinetikz}{
    \tikz[baseline=-0.5ex]\draw[black, solid, line width=1pt] (0,0) -- (0.5,0);
}
\DeclareRobustCommand{\inlinedashed}{
    \tikz[baseline=-0.5ex]\draw[black, dashed, line width=1pt] (0,0) -- (0.5,0);
}
\DeclareRobustCommand{\inlinedashdot}{
    \tikz[baseline=-0.5ex]\draw[black, dash dot dot, line width=1pt] (0,0) -- (0.5,0);
}
\begin{figure}[b!]
\begingroup
\begin{tabular}{c}
\includegraphics[width=1.0\linewidth]{st_t.eps} \\  
\end{tabular}
 \begin{tabular}{cc}
\includegraphics[width=0.24\textwidth]{st_t_m.eps} 
\includegraphics[width=0.235\textwidth]{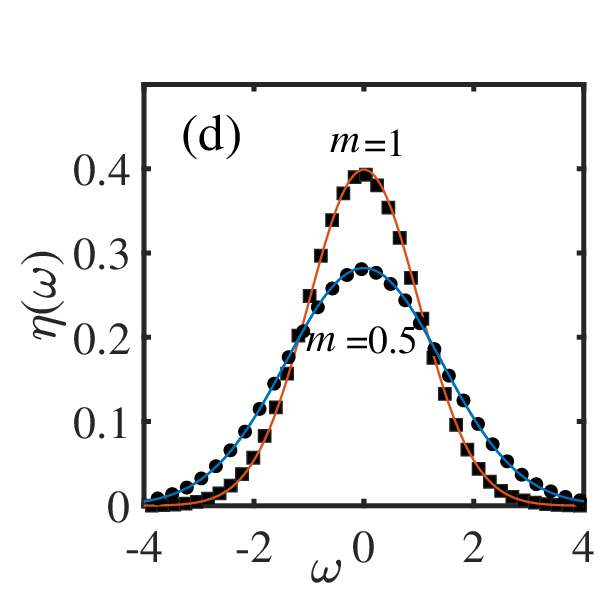} 
\end{tabular}
\endgroup
\caption{(Color online) (a) $s(t)$ vs $t$ for different values of $K_1=5$ (orange, \inlinetikz), 10 (violet, \inlinedashed), 15 (red, \inlinedashdot), at $K_2=5$. b) $s(t)$ vs $t$ for $K_1=10$ with different values of $K_2=0$ (violet, \inlinetikz) and 10 (orange, \inlinedashed ). (c) $s(t)$ vs $t$ for different values of $m$ with $D=1$ and $K_1=K_2=5$. (d) Distribution of frequency $\eta(\omega)$ plotted numerically (symbols) and analytically (solid lines) using Eq.~\ref{Sim_diff_Eq} and Eq.~\ref{dist_eta}, respectively.}
\label{st}
\end{figure}
Here, for $p=1$, $0\le r_1 \le 1$ quantifies the magnitude of global synchronization, and $\psi_1$ is the mean phase of all oscillators. $r_1=0$ implies a state in which all oscillators move incoherently around the circle, while $r_1=1$ indicates global synchronization. Moreover, the frequency order parameter is defined as follows,
\begin{equation}\nonumber \label{fre_order_par}
       {s e^{\iota\phi}=\frac{1}{N}\sum_{j=1}^{N}e^{\iota \omega_j}},
\end{equation}
where, $0\le s \le 1$ represents the magnitude of frequency synchronization. We express Eq.~\ref{Model_Eq} in the mean-field form by using the order parameter (Eq.~\ref{ord_par}) such that each oscillator interacts with the mean phase of all the other oscillators, 
\begin{equation}\label{Mean_field}
\begin{split}
    m\ddot{\theta_i}=-\dot{\theta_i}+&{\Omega_i}+{K_1 r_1}\sin(\psi_1-\theta_i)\\ &+{K_2 r_2 r_1}\sin(\psi_2-\psi_1-\theta_i)+\xi_i(t).
\end{split}
\end{equation}
Further, expressing Eq.~\ref{Mean_field} as a system of simultaneous first-order differential equations we get,
\begin{align}\label{Sim_diff_Eq}
    \dot{\theta_i} &= \omega_i, \nonumber \\ 
    \dot{\omega_i} &= \frac{1}{m}[-\omega_i + \Omega_i + K_1 r_1 \sin(\psi_1 - \theta_i) \nonumber\\
    & \quad + K_2 r_2 r_1 \sin(\psi_2 - \psi_1 - \theta_i) + \xi_i(t)].
\end{align}

\paragraph{\bf{Probability density function using Fokker-Plank equation:}}
 We employ the Fokker-Planck equation (FPE) to derive the probability density function $\rho(\theta,\omega,\Omega,t)$ for the oscillators,  
 \begin{figure}[t!]
    \centering
 \includegraphics[width=0.45\textwidth]{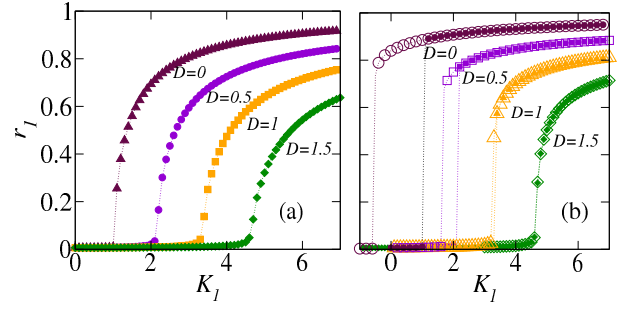}
    \caption{(Color online) $r_1$ vs $K_1$ at ($m=0.05$) illustrating shifts in $K_c$ with increasing the noise strength $D$. (a) $K_2=0$ and, (b) $K_2=5$. Results obtained numerically by simulating Eq.~\ref{Sim_diff_Eq} with Lorentzian frequency distribution ($\Omega_0=0 \,\,\& \,\,\Delta=0.5$). Open and filled symbols correspond to forward and backward numerical simulation predictions. }
    \label{low_inertia}
\end{figure}
\begin{equation}\label{FPE}
\begin{split}
\frac{\partial \rho}{\partial t}=\frac{D}{m^2}\frac{\partial^2 \rho}{\partial \omega^2}-\frac{1}{m} \frac{\partial}{\partial \omega}[(-\omega+\Omega+K_1r_1\sin(\psi_1-\theta)\\
K_2r_2r_1\sin(\psi_2-\psi_1-\theta))\rho]-\omega\frac{\partial \rho}{\partial \theta}.
\end{split}
\end{equation}
In the stationary state $\frac{\partial \rho}{\partial t}=0$, let us first consider that all oscillators are identical $g(\Omega)=\delta(\Omega)$. Further, the density distribution is $2\pi$ periodic in $\theta$ and decays as $\omega\rightarrow\pm\infty$. Following the normalization condition $\int_{0}^{2\pi}\int_{-\infty}^{\infty}\rho(\theta,\omega,\Omega,0) d\omega d\theta = 1 $. We look for the solution of the density function in the form of $\rho(\theta,\omega)= \eta(\omega)\chi(\theta)$, ultimately obtaining the expression for the distribution function of frequencies and phases.  Moreover, simulation of Eq.~\ref{Sim_diff_Eq} shows that the frequency order parameter is unaffected by $K_1$ and $K_2$ (Figs.~\ref{st}). Hence, solving Eq.~\ref{FPE} in the following manner, 
\begin{equation}\label{eta}\nonumber
    \frac{D}{m^2}\frac{d^2 \eta}{d\omega^2}+\frac{\omega}{m}\frac{d \eta}{d\omega}+\frac{\eta}{m}=0.
\end{equation}
\begin{equation}\label{chi}\nonumber
\begin{split}
   -\omega\eta\frac{d\chi}{d\theta}-\frac{1}{m}&\Bigl[K_1r_1\sin(\psi_1-\theta) +K_2r_2r_1 \\
  & \sin(\psi_2-\psi_1-\theta)\Bigr]\chi\frac{d\eta}{d\omega}=0.
\end{split}
\end{equation}
\begin{figure}[t!]
    \centering
\includegraphics[width=0.5\textwidth]{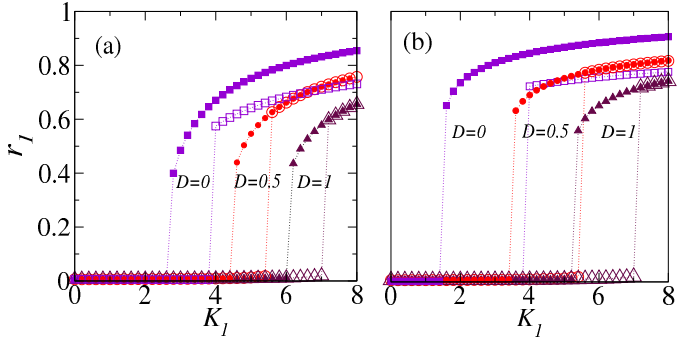}
    \caption{(Color online) $r_1$ as function of $K_1$ for $m=1$. (a) $K_2=0$, and (b) $K_2=5$ for different values of noise strength $D$. Here, $D=0$ (violet, square), $D=0.5$ (red, circle), $D=1$ (maroon, triangle). Results obtained by simulating Eq.~\ref{Sim_diff_Eq} with Lorentzian frequency distribution ($\Omega_0=0, \,\, \Delta=1$). Open and filled symbols correspond to forward and backward numerical simulation predictions, respectively.}
    \label{fig_m_1}
\end{figure}
By solving these differential equations, and applying boundary and normalization conditions, the distribution functions of frequencies and phases can be represented as follows,
\begin{equation}\label{dist_eta}
 \eta(\omega)=\sqrt\frac{m}{2\pi D}e^{\frac{-m\omega^2}{2D}}.
\end{equation}
\begin{equation}
\chi(\theta)=\frac{e^{K_1r_1\cos(\psi_1-\theta) +K_2r_2r_1\cos(\psi_2-\psi_1-\theta)}}{\int_{0}^{2\pi}e^{K_1r_1\cos(\psi_1-\theta) +K_2r_2r_1\cos(\psi_2-\psi_1-\theta)} d\theta}.
\end{equation}
Fig.~\ref{st}(d) illustrates the frequency distribution function derived analytically using Eq.~\ref{dist_eta} (solid lines), which match with the numerically obtained results from Eq.~\ref{Sim_diff_Eq} (symbols).

\paragraph{\bf{Perturbation analysis of incoherent state:}}
In the incoherent state,  phases are randomly distributed around the unit radius circle implying $r_1=r_2=0$.  Using Eq.~\ref{FPE} to obtain the density function of the incoherent state given as,
\begin{equation}
    \rho_0(\omega,\Omega)=\frac{1}{2\pi}\sqrt{\frac{m}{2\pi D}}e^{\frac{-m(\omega-\Omega)^2}{2D}}.
\end{equation}
Giving a small perturbation ($\epsilon << 1$) to the incoherent state, we get 
\begin{equation}\label{perturbation}
\rho(\theta,\omega,\Omega,t)=\rho_0(\omega,\Omega)+\epsilon\alpha(\theta,\omega,\Omega,t)+O(\epsilon^2).
\end{equation}
The order parameter  in the continuum limit can be written as,
\begin{equation}\label{cont_order_par}
\begin{split}
    r_1e^{\iota\psi_1}&=\int_{0}^{2\pi}\int_{-\infty}^{\infty}\int_{-\infty}^{\infty} e^{\iota\phi} \rho(\phi,\omega,\Omega,t)g(\Omega)d\Omega d\omega d\phi.\\
     r_2e^{\iota\psi_2}&=\int_{0}^{2\pi}\int_{-\infty}^{\infty}\int_{-\infty}^{\infty} e^{2\iota\phi} \rho(\phi,\omega,\Omega,t)g(\Omega)d\Omega d\omega d\phi.
\end{split}
\end{equation}
 Further, incorporating Eqs.(~\ref{perturbation} and  ~\ref{cont_order_par}) into Eq.~\ref{FPE}, and equating like terms in $\epsilon$ we obtain
\begin{widetext}
 \begin{equation}
    \frac{\partial \alpha}{\partial t}+\omega\frac{\partial \alpha}{\partial \theta}-\frac{D}{m^2}\frac{\partial^2 \alpha}{\partial \omega^2}-\frac{1}{m}\frac{\partial}{\partial \omega}\left((\omega-\Omega)\alpha\right)=
    -\frac{K_1}{m}\frac{\partial \rho_0}{\partial \omega}\int_{0}^{2\pi}\int_{-\infty}^{\infty}\int_{-\infty}^{\infty} \sin{(\phi-\theta)} \alpha(\phi,\omega,\Omega,t)g(\Omega)d\Omega d\omega d\phi.
\end{equation}
\end{widetext}
This equation is exactly the same as in Ref.~\cite{acebron2000} (Eq.~9), and  therefore the subsequent derivations are also the same as in Ref.~\cite{acebron2000}. For Lorentzian frequency distribution $g(\Omega)=\frac{\Delta}{\pi[(\Omega-\Omega_0)^2+\Delta^2]}$ with mean $\Omega_0=0$ and standard deviation $\Delta$, we get $K_c=2\Delta(m\Delta+1)+\frac{2(2+3m\Delta)}{2+m\Delta}D+O(D^2)$ for the forward transition point where oscillators jump from an incoherent to a synchronized state, applicable when noise strength is small ($D\ll 1$). For a negligible inertia $m$, $K_c=2(D+\Delta)$ is similar to the case of the Kuramoto Model with Gaussian white noise \cite{Strogatz91}. We find that $K_c$ is independent of $K_2$.
\begin{figure}[t!]
\begingroup
\begin{tabular}{cc}
\includegraphics[width=0.255\textwidth]{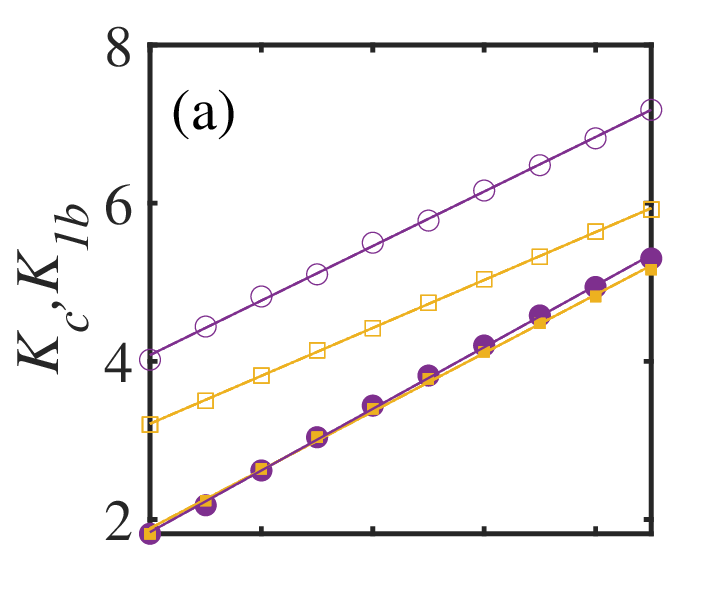}
 \includegraphics[width=0.24\textwidth]{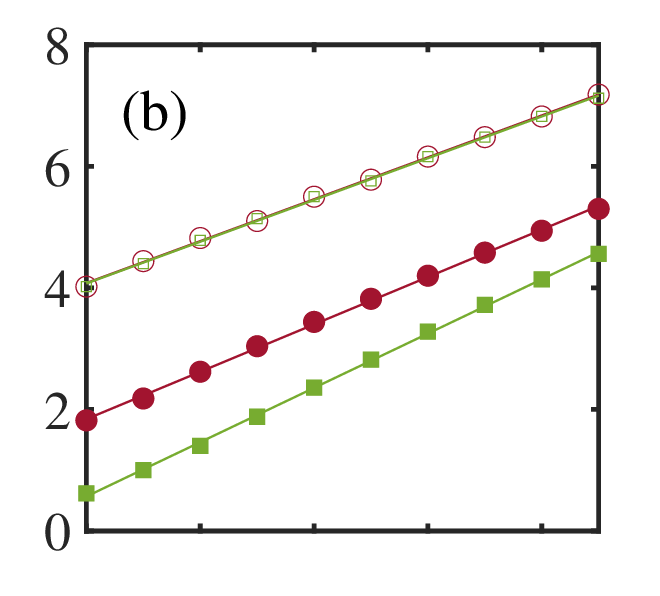}
 \vspace{-0.4cm}   
\end{tabular}
 \begin{tabular}{cc} \includegraphics[width=0.255\textwidth]{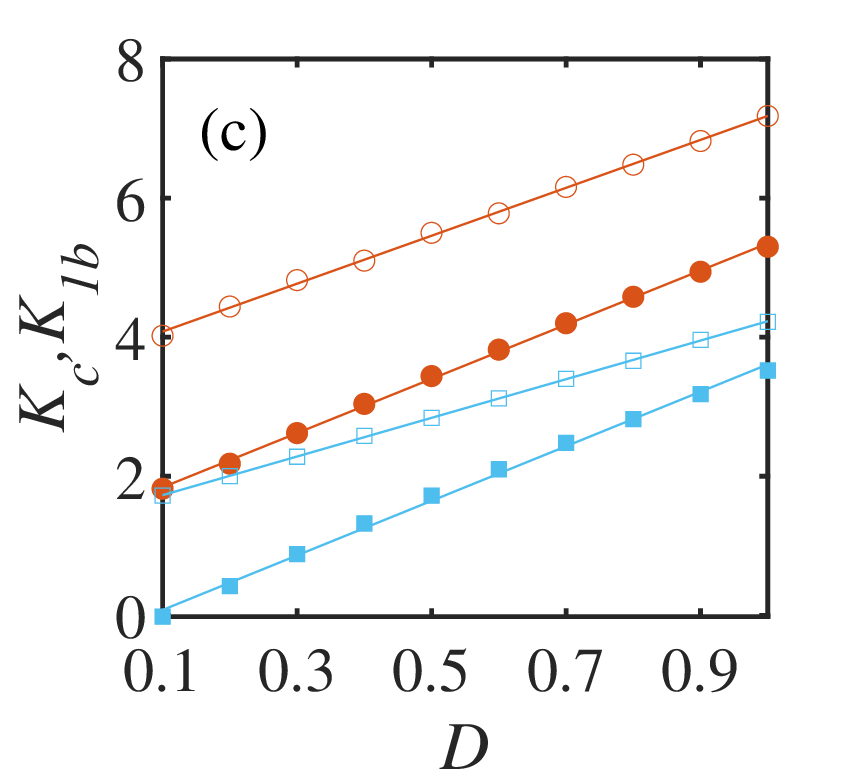} 
\includegraphics[width=0.245\textwidth]{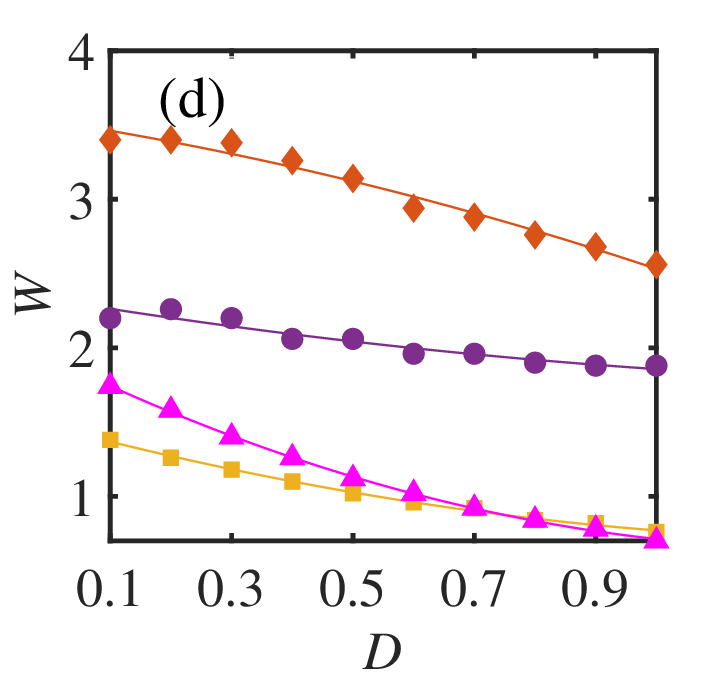} 
\end{tabular}
\endgroup
\caption{(Color online) Analysis of critical points and hysteresis width as functions of noise strength $D$. (a) Forward $K_c$ and backward $K_{1b}$ critical points for $K_2 = 5$ at different inertia values: $m=0.5$ (golden, square) and $m=1$ (violet, circle).   (b) Transition points for $m=1$ with varying $K_2$ values: $K_2 = 5$ (maroon, circle) and $K_2 = 8$ (green, square). (c) For $m=1$: $\Delta=0.5$ (sky blue, square) and $\Delta=1$ (orange, circle). (d) Hysteresis width $W$ vs $D$. Here,  $(m=0.5,\,\, K_2=5,\,\, \Delta=1)$ (golden, square), $(m=1,\,\, K_2=5,\,\, \Delta=0.5)$ (magenta, triangle), $(m=1,\,\, K_2=5,\,\, \Delta=1)$ (violet, circle), and $(m=1,\,\, K_2=8,\,\, \Delta=1)$ (orange, diamond). These results obtain numerically for $N=3\times10^4$ using Eq.~\ref{Sim_diff_Eq}. For (a), (b), and (c), open and filled symbols represent the forward and backward directions, respectively. Solid lines represent the fitted curve lines.}
\label{D_Kc}
\end{figure}
 \paragraph{\bf{Numerical Results:} }
Numerical simulations are performed using the Euler method with a time step $dt=0.05$, for Eq.~\ref{Model_Eq} by converting it to mean-field equation and then simultaneous first-order differential equations Eq.~\ref{Sim_diff_Eq}. Results are obtained for $N=20,000$ and $r_1$ averaged over $8\times 10^4$ iterations after removing an initial transient by changing $K_1$ in the forward and backward directions.

Fig.~\ref{low_inertia} depicts the phase transition for low  $m$ value indicating that critical transition point $K_c$ shifts towards higher positive coupling value with an increase in $D$. Fig.~\ref{low_inertia}(a) represents the second-order phase transition in absence of higher-order interactions ($K_2=0$). We consider the Lorentzian frequency distribution with standard deviation $\Delta=0.5$, so $K_c=2D+1$ correctly determines the transition point, for ($D<1$). Furthermore, Fig.~\ref{low_inertia}(b) for $K_2=5$ demonstrates that phase transition changes from first-order to second-order as we increase $D$, despite the presence of higher-order interactions. 

For $m=1$, Fig.~\ref{fig_m_1} (a, b)  depicts the first-order phase transition. With an increase in $D$, $K_c$ shifts towards higher value. Also, the backward transition point $K_{1b}$ at which the oscillators jump from synchronized to incoherent states shifts towards higher value. Further, Fig.~\ref{fig_m_1} (b) depicts that $K_c$ is not affected by the presence of $2-$simplex interactions term, as yielded by the perturbation analysis of the incoherent state that expression of $K_c$ comes out to be independent of $K_2$. Since $K_2$ impacts $K_{1b}$, thereby leading to an increase in the hysteresis width compared to the system with only pairwise interactions. 

Fig.~\ref{D_Kc} analyzes forward $K_c$ and backward $K_{1b}$ transition points as a function  of noise strength $D$, and
examines the impact of changes in $m$, $K_2$, $\Delta$. Fig.~\ref{D_Kc}(a) depicts that an increase in inertia $m$ shifts $K_c$ towards higher values, while $K_{1b}$ remains unaffected thereby increasing the hysteresis width. Fig.~\ref{D_Kc}(b) demonstrates that increasing $K_2$ leads to a shift in $K_{1b}$ towards {lower} values with $K_{c}$ remaining the same. This also results in an increased hysteresis width.  Fig.~\ref{D_Kc}(c) illustrates that an increase in $\Delta$ value shifts both $K_c$ and $K_{1b}$ towards higher values. As reflected by Fig.~\ref{D_Kc}(d) hysteresis width $W=K_c-K_{1b}$ decreases as $D$ increases. 

\paragraph{\bf{Overdamped system:} }
In overdamped systems, where the inertia $m$ is negligible, the Ott-Antonsen approach \cite{ott2008low} facilitates the dimensional reduction for intrinsic frequency drawn from the Lorentzian distribution as:
\begin{equation}\label{r_1}
 \dot{r_1}=-(D+\Delta)r_1+\frac{K_1(r_1-r_1^3)}{2}+\frac{K_2(r_1^3-r_1^5)}{2}.
\end{equation}
\begin{figure}
    \centering
\includegraphics[width=0.5\textwidth]{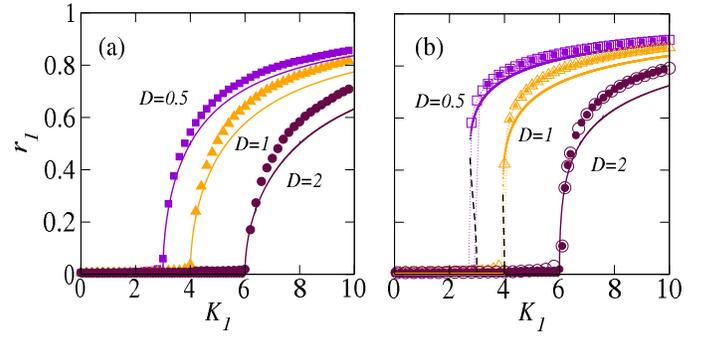}
    \caption{(Color online) $r_1$ as a function of $K_1$ for different values of $D$ and $K_2$. (a) $K_2=0$ and (b) $K_2=5$ depict shifts in $K_c$. These results are obtained numerically (Eq.~\ref{Mean_field}) for $m=0$, with Lorentzian frequency distribution ($\Omega_0=0, \,\, \Delta=1$). Forward and backward directions are represented by filled and open symbols, respectively. Analytical results (Eq.~\ref{r_1}) are illustrated with solid (stable) and dashed lines (unstable).}
    \label{KM}
\end{figure}
Eq.~\ref{r_1} is derived by equating the coefficient of $e^{\iota\theta}$ in the context of pairwise interactions only (as discussed in \cite{Sinha2023}). We extend the analysis to include $2$-simplex interactions. The forward critical transition point $K_c=2(D+\Delta)$, is a threshold where oscillators jump from an incoherent to a synchronized state, and for $K_1>K_c$, $r_1=0$ solution becomes unstable. By equating the coefficients of $e^{\iota n \theta}$, where $n \in \mathbb{N}$, we get $\dot{r_1}=-(nD+\Delta)r_1+\frac{K_1(r_1-r_1^3)}{2}+\frac{K_2(r_1^3-r_1^5)}{2}$. As we can see that for the noisy case $n$ arrives explicitly in $\dot{r_1}$ expression, hence Ott-Antonsen approach does not result in the dimensional reduction of  Eq.~\ref{Model_Eq} for overdamped system. However, by substituting $n=1$ in $\dot{r_1}$, the analytical predictions closely matches with those achieved through numerical simulations (Fig.~\ref{KM}). 

Fig.~\ref{KM} (a) $K_2=0$ shows second-order phase transition, and (b) $K_2=5$ manifests first-order phase transition. Increasing $D$ results in decrease in the hysteresis width, finally yielding second-order phase transition.

\paragraph{\bf{Conclusion:}} This study investigates noise-induced phase transitions in the Kuramoto model with inertia and $2$-simplex interactions. First, using Fokker-Plank equation, we determine the distribution function of frequency and phase for identical oscillators, demonstrating that the frequency order parameter while depends on inertia $m$, remains unaffected by coupling strengths $K_1$ and $K_2$. Our analysis further reveals that noise strength is a critical factor governing both forward and backward transition points. Using the perturbation analysis for Lorentzian frequency distribution, we show that the forward transition point is independent of the $2$-simplex interaction term and rather depends on inertia and noise strength. By employing the Ott-Antonsen approach we derive approximate analytical predictions for the overdamped system. A significant finding is that as noise strength increases in the overdamped system, the phase transition shifts from the first-order to the second-order even in the presence of higher-order interactions. These results underscore the interplay between noise, inertia, and higher-order interactions in synchronized systems. 

We hope that understanding these dynamical behaviours may be useful for studying  stability of power grid systems. By incorporating noise and higher-order interactions into the Kuramoto model, we can more accurately predict critical transition points which are essential for preventing systemic failures.
This improved modeling approach can aid in  designing of more robust power grid systems, capable of maintaining stability under perturbations and noise conditions. Furthermore, few direct possible extensions of this work are to have more realistic noise models \cite{bag2007} and incorporation of multiplicative noise \cite{pinto2017}.

\paragraph*{Acknowledgement:}
SJ and PR acknowledge Govt of India SERB Power grant SPF/2021/000136 and PMRF grant PMRF/2023/2103358, respectively.

\end{document}